\documentclass[
    reprint,
    amsmath,
    amssymb,
    aps,
    prl,
    superscriptaddress
]{revtex4-2}
\usepackage[colorlinks=true]{hyperref}
\usepackage{tabularx}
\usepackage{graphicx}
\usepackage{braket}
\usepackage{mathtools}
\usepackage{bm}
\newcommand{\Tr}{\mathrm{Tr}}
\newcommand{\im}{\mathrm{i}}
\newcommand{\e}{\mathrm{e}}

\newcommand{\calD}{\mathcal{D}}
\newcommand{\calU}{\mathcal{U}}
\newcommand{\calW}{\mathcal{W}}
\newcommand{\calS}{\mathcal{S}}

\begin{document}

\title{Realization of quantum signal processing on a noisy quantum computer}

\author{Yuta Kikuchi}
\email{yuta.kikuchi@quantinuum.com}
\affiliation{Quantinuum K.K., Otemachi Financial City Grand Cube 3F, 1-9-2 Otemachi, Chiyoda-ku, Tokyo, Japan}
\affiliation{Interdisciplinary Theoretical and Mathematical Sciences Program (iTHEMS), RIKEN, Wako, Saitama 351-0198, Japan}

\author{Conor Mc Keever}
\affiliation{Quantinuum, Partnership House, Carlisle Place, London SW1P 1BX, United Kingdom}

\author{Luuk Coopmans}
\affiliation{Quantinuum, Partnership House, Carlisle Place, London SW1P 1BX, United Kingdom}

\author{Michael Lubasch}
\affiliation{Quantinuum, Partnership House, Carlisle Place, London SW1P 1BX, United Kingdom}

\author{Marcello Benedetti}
\email{marcello.benedetti@quantinuum.com}
\affiliation{Quantinuum, Partnership House, Carlisle Place, London SW1P 1BX, United Kingdom}

\date{September 27, 2023}

\begin{abstract}
Quantum signal processing (QSP) is a powerful toolbox for the design of quantum algorithms and can lead to asymptotically optimal computational costs.
Its realization on noisy quantum computers without fault tolerance, however, is challenging because it requires a deep quantum circuit in general.
We propose a strategy to run an entire QSP protocol on noisy quantum hardware by carefully reducing overhead costs at each step.
To illustrate the approach, we consider the application of Hamiltonian simulation for which QSP implements a polynomial approximation of the time evolution operator.
We test the protocol by running the algorithm on the Quantinuum H1-1 trapped-ion quantum computer powered by Honeywell.
In particular, we compute the time dependence of bipartite entanglement entropies for Ising spin chains and find good agreements with exact numerical simulations.
To make the best use of the device, we determine optimal experimental parameters by using a simplified error model for the hardware and numerically studying the trade-off between Hamiltonian simulation time, polynomial degree, and total accuracy.
Our results are the first step in the experimental realization of QSP-based quantum algorithms.
\end{abstract}

\maketitle

\section{Introduction}

Several quantum algorithms are known to outperform their classical counterparts by computational costs that asymptotically scale better, e.g.,\ Shor's prime factoring algorithm~\cite{Sh94}, Hamiltonian simulation~\cite{Lloyd1996, Ll97} and Grover search~\cite{Gr98, BrEtAl02}.
Their realization on actual quantum computers, however, requires additional qubits and gates to correct errors that naturally occur in real physical devices.
Currently available noisy quantum computers are not capable yet of running such quantum algorithms for large problem sizes.

In the context of noisy quantum circuits, there are two regimes in which the classical computational requirements for simulating a quantum computer remain tractable. First, shallow circuits typically generate small amounts of entanglement making them amenable to classical simulation. Second, deep circuits quickly accumulate errors causing decoherence towards a regime which can also be treated efficiently on classical computers~\cite{Zhou2020what, StFr21}.
Between these two extremes, there is an optimal working point at which maximum non-trivial quantum correlation is attained and where accurate simulation may become challenging for a classical computer~\cite{Noh2020efficientclassical}. In light of this, a promising route towards achieving a genuine quantum advantage without fault tolerance is to realize the aforementioned algorithms while operating the computer at its optimal working point. In order to design such an algorithm, it is therefore essential to account for the influence of noise on the circuits which implement it.

In this work, we propose to heuristically optimize the depth of quantum circuits and operate where we can make the most out of our noisy quantum computer. With this heuristic approach, we provide the first realization of quantum signal processing (QSP) on a trapped-ion quantum computer. 
QSP was proposed in~\cite{Low_2016} and is now recognized as one of the most powerful frameworks for developing quantum algorithms. It gives a unifying perspective on seemingly distinct algorithms such as amplitude amplification and the quantum linear systems algorithm and improves on their computational resources~\cite{Gilyen_2019,Martyn_2021_grand}.
Such flexibility stems from the fact that QSP allows one to apply almost any polynomial transformation to an input scalar or matrix. In the literature, QSP often refers to a polynomial transformation applied to an input scalar, and its generalizations apply a polynomial transformation to eigenvalues (QET) or singular values (QSVT) of an input matrix. Throughout this article, we do not make such a distinction and refer to all these protocols as QSP.

Hamiltonian simulation is an example where QSP provides an improved asymptotic scaling over other algorithms.
Since Feynman's seminal proposal~\cite{Feynman1982}, Hamiltonian simulation has been a fundamental problem of quantum computing. 
An efficient Hamiltonian simulation algorithm allows us to simulate the real-time dynamics of a quantum system described by a Hamiltonian $H$ with computational resources scaling at most polynomially in evolution time $t$, system size $n$, and inverse of required accuracy $1/\epsilon$.
Extensive studies have been devoted to exploring efficient algorithms for Hamiltonian simulation, which include product formulas~\cite{Lloyd1996,Suzuki1991,berry2007efficient,Childs2021}, quantum walks~\cite{berry2009black}, the truncated Taylor-series expansion~\cite{Berry2015Taylor}, randomized protocols~\cite{Poulin2011,Childs2019fasterquantum,Campbell2019,Chen2021,Zhao2022}, and making use of classical optimization techniques~\cite{Tepaske_2022, McKeever_2022, Mansuroglu_2023}.
Nowadays, the QSP-based algorithm is known to exhibit nearly optimal asymptotic scaling~\cite{Low_2017,Gilyen_2019,Low2019hamiltonian} (see also~\cite{Childs2018} for a comparative survey).

\begin{figure*}
    \centering
    \includegraphics[width=\textwidth]{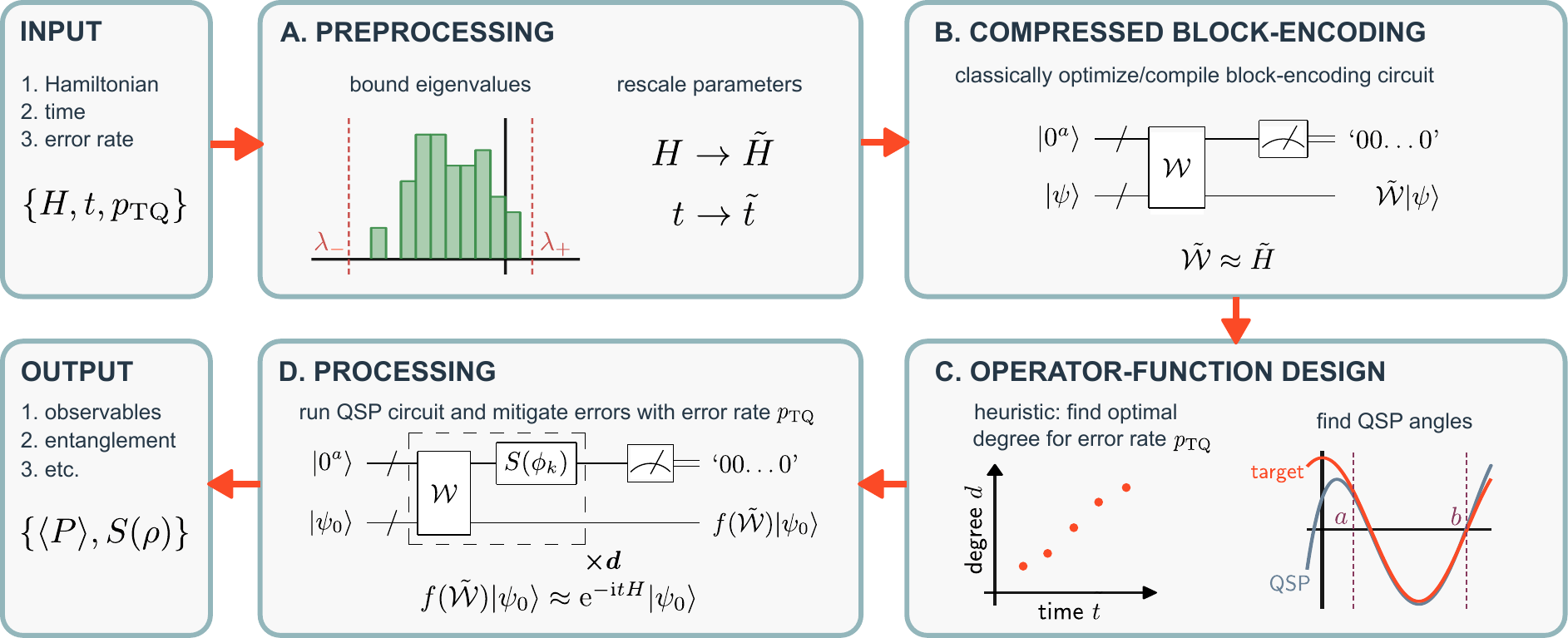}
    \caption{The proposed protocol for the realization of QSP on a noisy quantum computer. We choose Hamiltonian simulation as the application. We start with a necessary preprocessing step \hyperref[sec:preprocessing]{(A)} that maps the input parameters to an effective Hamiltonian $\tilde{H}$ and an effective simulation time $\tilde{t}$. In step \hyperref[sec:compressBE]{(B)}, $\tilde{H}$ is embedded in a unitary operator. By classically optimizing/compiling a circuit $\calW$ this step produces a compressed version of a block-encoding circuit. Next, in the operator-function design \hyperref[sec:design]{(C)}, we approximate the real-time evolution function, $\e^{-\im xt}$, by a polynomial $f(x)$ of degree $d$. While increasing the degree leads to a more accurate polynomial approximation, the computation suffers from larger noise effects. This is due to the growing depth of the QSP circuit, consisting of $\mathcal{O}(d)$ primitive gates. By accounting for the error rate $p_\mathrm{TQ}$ of two-qubit gates, we heuristically estimate the optimal degree yielding the smallest combined error. The processing step \hyperref[sec:processing]{(D)} finally realizes QSP using the compressed block-encoding circuit $\calW$ and the designed polynomial $f(x)$. Upon postselection on the ancilla's measurement outcomes, we obtain an approximation to the desired real-time evolution $\e^{-\im H t}$. An error mitigation scheme based on the error rate $p_\mathrm{TQ}$ further reduces the effect of noise on the output.}
    \label{fig:protocol}
\end{figure*}

In~\cite{Dong2022}, the authors demonstrate the QSP protocol using random Hamiltonians on a superconducting device for the purpose of benchmarking.
The present work takes a step forward by realizing QSP on the Quantinuum H1-1 trapped-ion quantum computer and performing the Hamiltonian simulation of physically relevant quantum systems. After the release of the present manuscript, another group demonstrated QSP for the task of quantum channel discrimination~\cite{Debry2023experimental}.
 
\section{Results}
\subsection{Review of Hamiltonian simulation by quantum signal processing}
\label{sec:review}

The Hamiltonian simulation algorithm solves the real-time dynamics of a quantum system by applying a real-time evolution operator $\e^{-\im H t}$ to some initial state $\ket{\psi_0}$, where the Hamiltonian $H$ is given by a Hermitian operator in this work. We employ QSP in order to find an approximate real-time evolution operator that can be efficiently implemented on a quantum computer. 
QSP outputs a degree-$d$ polynomial ${f\in\mathbb{C}[x]}$ using a sequence of unitary operators~\cite{Low_2016,Low_2017,Gilyen_2019,Low2019hamiltonian},
\begin{align}
\label{eq:qsp}
    U_{\mathrm{QSP}}&:=
    \prod_{k=1}^{d} \big[S(\phi_k)W(x)\big]
    =
    \begin{pmatrix}
        f(x) & \ast \\
        \ast & \ast
    \end{pmatrix},
    \\
    S(\phi) &:=
    \begin{pmatrix}
        \e^{\im\phi} & 0 \\
        0 & \e^{-\im\phi}
    \end{pmatrix},
    \\
    W(x) &:= 
    \begin{pmatrix}
        x & \sqrt{1-x^2} \\
        \sqrt{1-x^2} & -x
    \end{pmatrix},
\end{align}
where $\ast$ stands for an unspecified entry.
Here, we follow the convention of Corollary~8 in~\cite{Gilyen_2019} (preprint version), where $W(x)$ takes the form of a reflection operator. For a polynomial $f(x)$ that satisfies certain conditions~\cite{Gilyen_2019,Martyn_2021_grand} there always exists a set of QSP angles $\{\phi_k\}$. The conditions are: (i) $f$ must have parity-$(d \mathrm{\ mod\ }2)$, (ii) $|f(x)|\le1$ for all $x\in[-1,1]$, (iii) $|f(x)|\ge1$ for all $x\in(-\infty,1]\cup(1,\infty]$, and (iv) $f(\im x)f^*(\im x)\ge1$ for all $x\in\mathbb{R}$ if $d$ is even. The function $f(x)$ is implemented by computing such angles $\{\phi_k\}$, and is encoded in the expectation $\bra{0}U_\text{QSP}\ket{0}$.
It is evident from Eq.~\eqref{eq:qsp} that the circuit depth is proportional to the degree $d$. 

Finding an efficient Hamiltonian simulation algorithm with QSP starts by approximating the function $\e^{-\im xt}$ with a fixed-degree polynomial on an interval $I \subseteq [-1,1]$.
Given time $t > 0$ and accuracy $\epsilon_\mathrm{poly}$, we find a polynomial $f$ such that
\begin{align}
\label{eq:poly_err}
    \max_{x\in I} | f(x) - \e^{-\im xt} | \leq \epsilon_\mathrm{poly} .
\end{align}
One way to find $f$ is to consider the polynomial approximation to the exponential function given by the Jacobi-Anger expansion~\cite{Low_2017},
\begin{align}
\begin{split}
    &\e^{-\im xt} = \cos(xt) - \im \sin(xt),
    \\
    &\cos(xt)=J_0(t) + 2\sum_{k=1}^{\infty}J_{2k}(t)T_{2k}(x),
    \\
    &\sin(xt)=2\sum_{k=1}^{\infty}J_{2k+1}(t)T_{2k+1}(x),
\end{split}
\end{align}
where $J_i(t)$ is a Bessel function of order $i$, and $T_i(x)$ is a Chebyshev polynomial of order $i$. Tolerating an error $\epsilon_\mathrm{poly}$, the polynomial can be truncated at degree
\begin{align}
\label{eq:degree_poly}
    d = \Theta\left(t + \frac{\log(1/\epsilon_\mathrm{poly})}{\log(\e + \log(1/\epsilon_\mathrm{poly})/t)}\right),
\end{align}
which is almost linear in $t$ and logarithmic in $1/\epsilon_\mathrm{poly}$. Here, we use the big-$\Theta$ notation, i.e., for functions $f$ and $g$ we write $f(x)=\Theta(g(x))$ if there exist constants $c_1$, $c_2$, and $x_0$ such that $c_1g(x)\le f(x)\le c_2g(x)$ for any $x>x_0$.

The goal is to apply this polynomial transformation to the eigenvalues of the Hamiltonian $H$. This is achieved by block encoding $H$, i.e., embedding $H$ in a unitary operator $\calW(H)$ acting on a larger Hilbert space. A number of block-encoding methods have been proposed in the literature~\cite{Gilyen_2019,Low2019hamiltonian,Chakraborty2019power,Camps:2022jnx,Camps:2022uyk} and their applicability depends on the form of the Hamiltonian. For instance, one can employ the linear-combination-of-unitary (LCU) method when $H$ is given as a weighted sum of unitary operators~\cite{Childs2012}.
Then, by identifying a subspace analogous to a one-qubit space, the block-encoding unitary $\calW(H)$ and a generalized rotation operator $\calS(\phi)$ behave like the single-qubit operations $W(x)$ and $S(\phi)$ in Eq.~\eqref{eq:qsp}.

Our aim is to run a small-scale QSP-based Hamiltonian simulation on a quantum computer with no fault-tolerance mechanism. This is challenging because noise limits the maximum depth of our circuits. 
We present a practical protocol to run the Hamiltonian simulation by QSP, while taking hardware noise into account.

\subsection{Preprocessing}
\label{sec:preprocessing}

Recall that QSP applies a polynomial transformation to the eigenvalues of the Hamiltonian. The eigenvalues need to be rescaled in a suitable interval so that the Hamiltonian can be encoded as a sub-block of a unitary operator. By unitarity, the largest possible interval in Eq.~\eqref{eq:poly_err} is $[-1,1]$. However, the protocol is made more efficient if we further narrow the interval down to $[0,1]$ and approximate $\e^{-\im xt}$ by an even function of $x$~\cite{Martyn2021efficient}. A general preprocessing method to rescale the spectrum of $H$ in $[a, b]\subseteq[0,1]$ is given by
\begin{equation}
\label{eq:rescaling_H}
    \tilde{H} = \frac{(H - \lambda_{-} I)(b - a)}{\lambda_{+} - \lambda_{-}} + a I ,
\end{equation}
where $\lambda_{+}$ and $\lambda_{-}$ are upper and lower bounds on the eigenvalues, respectively (see Fig.~\ref{fig:protocol}A). To recover the desired time evolution, we counterbalance with a time rescaling
\begin{equation}
\label{eq:rescaling_t}
    \tilde{t} = \frac{t (\lambda_{+} - \lambda_{-}) }{ b - a} .
\end{equation}
This yields the desired real-time evolution operator up to an irrelevant global phase: $\e^{-\im \tilde{t} \tilde{H}} = \e^{-\im \phi} \e^{-\im t H}$, where $\phi = t(a \lambda_{+}  - b\lambda_{-})/(b-a)$. The exact minimum $\lambda_\text{min}$ and maximum $\lambda_\text{max}$ eigenvalues are unknown and finding them is computationally intractable in general~\cite{kitaev2002classical,Kempe_2003,Kempe_2006}. That is why we resort to bounds. Equation~\eqref{eq:rescaling_t} shows that the effective evolution time $\tilde{t}$ increases as the QSP interval $[a, b]$ gets smaller, and as the eigenvalue bounds get looser. 
For example, suppose $\lambda_\pm$ are taken such that $(\lambda_{+}-\lambda_\text{max})/|\lambda_\text{max}|=(\lambda_\text{min}-\lambda_{-})/|\lambda_\text{min}|=r\ge0$, i.e., the bounds $\lambda_{+/-}$ are $100r$\% off from $\lambda_\mathrm{max/min}$. From Eq.~\eqref{eq:rescaling_t} we obtain
\begin{equation}
\label{eq:effective_t}
    \tilde{t} = \frac{t (\lambda_\text{max} - \lambda_\text{min}) }{b - a} +  \frac{ r t (|\lambda_\text{max}| + |\lambda_\text{min}|)}{b-a} .
\end{equation}
The first term is the smallest effective time achievable, while the second term is extra overhead. Note that $\tilde{t}$ determines the polynomial degree $d$ (e.g., Eq.~\eqref{eq:degree_poly} for the truncated Jacobi-Anger expansion), and thus the circuit depth.

When the Hamiltonian is provided as a weighted sum $H = \sum_k c_k H_k$ of operators $\{H_k\}$, simple bounds are readily available: $\lambda_{\pm} = \pm \sum_k |c_k| \; \lVert{H_k}\rVert$, where $\lVert\cdot\rVert$ is the spectral norm. Tighter bounds can be obtained by relaxing the ground-state constraints~\cite{Baumgratz2012lower, BaHu12} and/or exploiting some structure in the Hamiltonian. For translation-invariant systems, the Anderson bound~\cite{Anderson1951}, and a particular semi-definite program relaxation, can provide a lower bound with an error that is independent of system size~\cite{Eisert2023}. Furthermore, for a large class of local Hamiltonians, one can formulate a hierarchy of semi-definite programming constraints with increasing complexity that can be solved numerically with tensor network and renormalization group techniques~\cite{Kull2022}. 

\subsection{Compressed block-encoding} 
\label{sec:compressBE}

The second key step of the protocol (Fig.~\ref{fig:protocol}B) is to input the Hamiltonian to the quantum computer so that it can be processed. For $\epsilon_\mathrm{BE} \geq 0$, a block-encoding $\calW$ of $\tilde{H}$ is defined by
\begin{align}
\label{eq:block_enc_def}
\begin{split}
	&\left\| \tilde{\calW} - \tilde{H} \right\|_\mathrm{F}
    =\epsilon_\mathrm{BE}, \\
    &\tilde{\calW} := (\bra{0^a}\otimes I)\calW(\ket{0^a}\otimes I),
\end{split}
\end{align}
where $\| \cdot \|_\mathrm{F}$ is the Frobenius norm and the integer $a$ is the number of ancillary qubits. Note that $(\bra{0^a}\otimes I) \cdot (\ket{0^a}\otimes I)$ projects onto the subspace where the ancillary qubits are in the all-zero state.
The accuracy of the block encoding is specified by the parameter $\epsilon_\mathrm{BE}$.

Depending on the form of $\tilde{H}$, there exist different block-encoding methods~\cite{Gilyen_2019,Low2019hamiltonian,Chakraborty2019power,Camps:2022jnx,Camps:2022uyk,Childs2012}. While such generic methods are scalable in principle, the required number of ancillary qubits and the circuit depth may preclude an implementation on current noisy quantum devices. Here, we propose two ways to overcome this by compressing the block-encoding circuit. 

First, we use a parameterized quantum circuit $\cal{W}=\calW(\bm{\theta})$ as ansatz and minimize Eq.~\eqref{eq:block_enc_def} with respect to the parameters $\bm{\theta}$. The possible presence of barren plateaus in the optimization landscape could prohibit quantum-classical hybrid methods from being efficient at larger system sizes~\cite{McClean_2018, Grant2019initialization, Cerezo_2021}. In this case, a fully classical approach is preferable~\cite{Martin_2022}. We thus suggest to use tensor network ans\"{a}tze that can be efficiently optimized on a classical computer.

Second, we make use of multiplexor circuit compilation to compress the LCU block-encoding circuit~\cite{Sivarajah_2021,Yao_to_appear}. The multiplexor compilation reduces the number of elementary gates required to implement sequential multi-controlled unitary operations which are heavily used in the LCU circuit. Since the compilation adopted here does not introduce approximation error, it provides an exact block-encoding, i.e., $\epsilon_\mathrm{BE}=0$.

In the Methods section we discuss both approaches in more detail.

\subsection{Operator-function design}
\label{sec:design}

\begin{figure*}
\centering
\includegraphics[width=\textwidth]{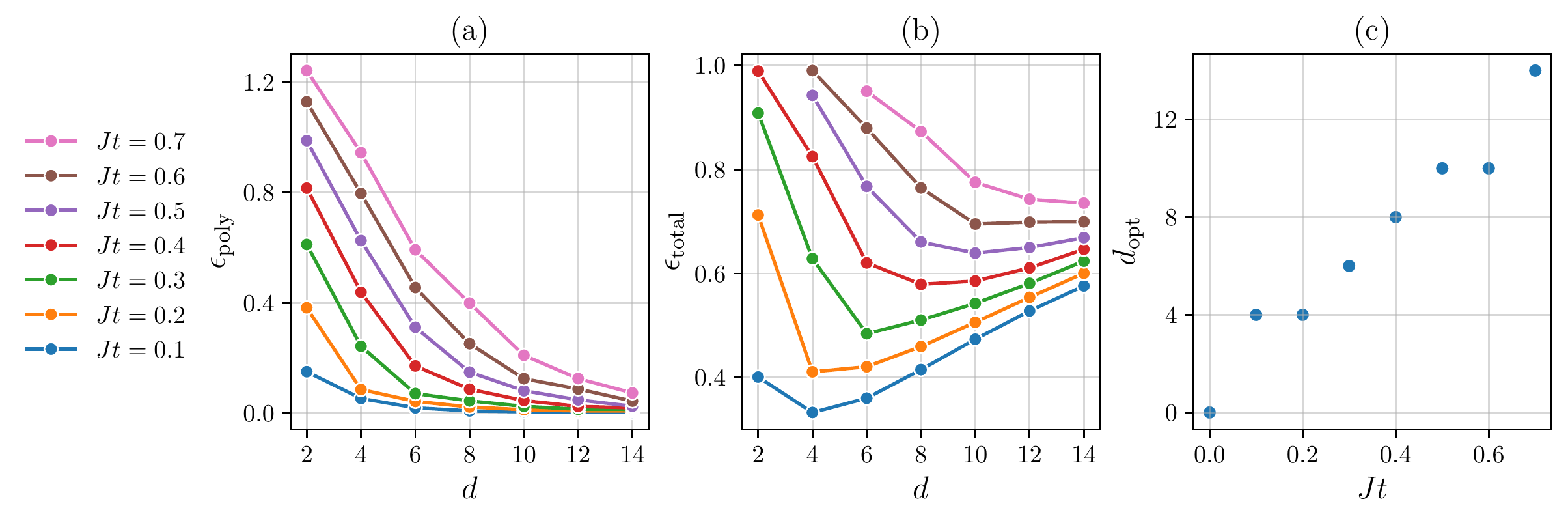}
\caption{Heuristic search of optimal parameters for the five-qubit hardware experiment. (a) Accuracy of QSP angle optimization, Eq.~\eqref{eq:find_angles}, using \texttt{pyqsp}~\cite{pyqsp}. (b) Upper bound to the infidelity, Eq.~\eqref{eq:total_infidelity}, as a function of degree $d$ and evolution time $Jt$.
(c) For each evolution time, the optimal degree $d_\mathrm{opt}$ is the degree that minimizes the total error $\epsilon_\mathrm{total}$ in (b).}
\label{fig:error_degree}
\end{figure*}

The depth of a QSP circuit is proportional to the degree $d$ of the polynomial. When using noisy devices, we must fix $d$ so that the final circuit has a reasonable fidelity. Later on, we provide a heuristic to choose $d$ as a function of $\tilde{t}$ and hardware noise. For now, let us assume that $d$ is fixed and proceed to the function design (Fig.~\ref{fig:protocol}C).
Instead of using the Jacobi-Anger expansion, we numerically optimize the QSP angles $\{\phi_k\}$. The preprocessing step has rescaled the eigenvalues of $H$ in $[a,b] \subseteq [0,1]$, so we restrict the optimization to that interval. Furthermore, we can utilize polynomials of even parity, i.e., QSP polynomials of even degree $d$. The resulting accuracy is
\begin{align}
\label{eq:find_angles}
    \epsilon_\mathrm{poly} 
    = \min_{\{\phi_k\}} \max_{x\in [a, b]} \big| \bra{0} U_{\mathrm{QSP}}(\{\phi_k\}) \ket{0} - \e^{-\im x \tilde{t}} \big|.
\end{align}

Figure~\ref{fig:error_degree}(a) shows the accuracy for different values of degree and evolution time. For each value of $d$, we find the QSP angle sequence using a dedicated python package called \texttt{pyqsp}~\cite{pyqsp}. As expected, the error decreases as the degree gets larger for a given evolution time. It is also observed that the error increases as the evolution time gets longer for a fixed degree.

The error stemming from both block-encoding~\eqref{eq:block_enc_def} and operator-function design~\eqref{eq:find_angles} propagates to the accuracy of the whole algorithm. 
This is found by expanding the error as~\cite{Gilyen_2019,Chakraborty2019power},
\begin{align}
\label{eq:qsp_epsilon}
\begin{split}
    &\big\lVert\e^{-\im \tilde{t} \tilde{H}} - f(\tilde{\calW}) \big\rVert
    \\
    &\le  \big\lVert\e^{-\im \tilde{t} \tilde{H}} -\e^{-\im \tilde{t} \tilde{\calW}}\big\rVert
    + \big\lVert \e^{-\im \tilde{t} \tilde{\calW}}- f(\tilde{\calW}) \big\rVert
    \\
    &\le |\tilde{t}| \; \big\lVert \tilde{H} - \tilde{\calW} \big\rVert_\mathrm{F}
    + \big\lVert \e^{-\im \tilde{t} \tilde{\calW}}- f(\tilde{\calW}) \big\rVert
    \\
    &\overset{\eqref{eq:block_enc_def},\eqref{eq:find_angles}}{=} |\tilde{t}| \; \epsilon_\mathrm{BE} + \epsilon_\mathrm{poly} =:\epsilon_\mathrm{QSP},
\end{split}
\end{align}
where we have defined $f(\tilde{\calW}):= \sum_{\lambda_{\tilde{\calW}}}f(\lambda_{\tilde{\calW}})\ket{\lambda_{\tilde{\calW}}}\bra{\lambda_{\tilde{\calW}}}$ with the eigenstates $\{\ket{\lambda_{\tilde{\calW}}}\}$ of $\tilde{\calW}$ such that $\tilde{\calW}\ket{\lambda_{\tilde{\calW}}}=\lambda_{\tilde{\calW}}\ket{\lambda_{\tilde{\calW}}}$.
In the third line, we use inequality $\lVert\e^{-\im \tilde{t} \tilde{H}} -\e^{-\im \tilde{t} \tilde{\calW}}\rVert\le|\tilde{t}| \; \lVert \tilde{H} - \tilde{\calW} \rVert_\mathrm{F}$ (see Lemma 50 in~\cite{Chakraborty2019power}, preprint version) and the fact that the spectral norm is upper bounded by the Frobenius norm.

Let us now incorporate the effect of hardware noise via a simple noise model. This allows us to develop a heuristic for estimating the optimal polynomial degree, given the evolution time and the noise rate of our quantum device.
Letting $\ket{\psi_0}$ be a $n$-qubit initial state and $\ket{0^a}$ be the $a$-qubit ancillary state, the quantum computation is described by 
\begin{align}
    \sigma = \calU_\mathrm{QSP}(\ket{0^a}\bra{0^a}\otimes\ket{\psi_0}\bra{\psi_0})\calU_\mathrm{QSP}^\dag,
\end{align}
where $\calU_\mathrm{QSP}$ represents the unitary implementing the QSP protocol, which will be defined later in Eq.~\eqref{eq:QSP_matrix}.
We model the noise effect of the hardware with the depolarizing channel $\calD_p$ acting on the entire system.
It alters the state to
\begin{align}
\label{eq:noisy_state}
    \calD_p[\sigma] 
    = (1-p)\sigma + p\frac{I}{2^{n+a}},
\end{align}
where we set $p=1 - (1-p_\mathrm{TQ})^{N_\mathrm{TQ}}$ with the two-qubit gate infidelity $p_\mathrm{TQ}$ and the number of two-qubit gates $N_\mathrm{TQ}$ in the $\calU_\mathrm{QSP}$ circuit.
The fidelity between this state and the ideal target state $\ket{\psi_{\tilde{t}}}:=\e^{-\im\tilde{H}\tilde{t}}\ket{\psi_0}$ quantifies the error,
\begin{align}
\begin{split}
\label{eq:noisy_fidelity}
    &(\bra{0^a}\otimes\bra{\psi_{\tilde{t}}}) \calD_p[\sigma] (\ket{0^a}\otimes\ket{\psi_{\tilde{t}}})
    \\
    &= (1-p) \big|\bra{\psi_{\tilde{t}}}f(\tilde{\calW})\ket{\psi_0} \big|^2
    +\frac{p}{2^{n+a}},
\end{split}
\end{align}
Thus, the corresponding infidelity is bounded as
\begin{align}
\label{eq:total_infidelity}
    & 1-
    (\bra{0^a}\otimes\bra{\psi_{\tilde{t}}})\calD_p[\sigma](\ket{0^a}\otimes\ket{\psi_{\tilde{t}}})
    \nonumber\\
	&= 1 - (1-p) \big| 1-\bra{\psi_{\tilde{t}}} \big(\e^{-\im\tilde{H}\tilde{t}}-f(\tilde{\calW}) \big)\ket{\psi_0} \big|^2 - \frac{p}{2^{n+a}}
    \nonumber\\
	&\overset{\eqref{eq:qsp_epsilon}}{\le} 
 1 - (1-p) (1-\epsilon_\mathrm{QSP})^2 - \frac{p}{2^{n+a}}
 =:\epsilon_\mathrm{total}.
\end{align} 

Figure~\ref{fig:error_degree}(b) shows the upper bound in Eq.~\eqref{eq:total_infidelity} as a function of degree and evolution time, where the algorithmic error $\epsilon_\mathrm{QSP}$ [Eq.~\eqref{eq:qsp_epsilon}] is obtained for the Hamiltonian given in Eq.~\eqref{eq:spin_hamiltonian}.
The two-qubit gate error rate is set to $p_\mathrm{TQ}=2.577\times10^{-3}$ (see Methods for details) and the circuits of degree $d\in\{2,4,6,8,10,12,14\}$ contain $N_\mathrm{TQ}\in\{52,98,144,190,236,282,328\}$ two-qubit gates, respectively.
In contrast to the operator-function design error in Fig.~\ref{fig:error_degree}(a), the total error in Fig.~\ref{fig:error_degree}(b) has a sweet spot for each value of $Jt$. 
Intuitively, the increase of the degree reduces the algorithmic error $\epsilon_\mathrm{QSP}$ while making the noise effect more prominent due to the larger circuit depth. This motivates the following heuristic: for a given evolution time, pick the degree that minimizes the upper bound on the total error~\eqref{eq:total_infidelity} (see \cite{Vrana_2014,Cohn_2016}, where a similar approach has been applied to Grover's algorithm). Importantly, this step of the protocol does not require the use of a quantum computer. The optimal degree for Eq.~\eqref{eq:total_infidelity} is found numerically using classical computation. Additionally, the sweet spot may coincide with the hardware's optimal working point where we expect a classical simulation of the corresponding noisy quantum circuit to be most challenging~\cite{Zhou2020what, Noh2020efficientclassical}, further justifying our heuristic choice. 

Figure~\ref{fig:error_degree}(c) shows that the optimal degree $d_\mathrm{opt}$ is approximately linear in the evolution time $t$. The estimated degrees are corroborated by the complementary numerical study that we carried out and presented in the Methods section. 
It is important to emphasize that our approximately linear scaling in time is different from the one expected by noiseless QSP. Our heuristic is designed to run the noisy quantum computer to its full potential, but may still produce large errors. This happens when the simulation parameters $\{H, t, p_\mathrm{TQ}\}$ are not compatible in the first place. For instance, at a fixed error rate $p_\mathrm{TQ}$ and large simulation time $t$, it is reasonable to expect a large infidelity. In contrast, Hamiltonian simulation by noiseless QSP achieves linear scaling in time while providing full control over the total error. For example, one can use a perfect block-encoding, $\epsilon_\mathrm{BE} = 0$, along with the desired approximation error $\epsilon_\mathrm{poly}$ in Eq.~\eqref{eq:degree_poly}.

\subsection{Processing}
\label{sec:processing}

In this last step of the protocol, we apply the polynomial $f$ found in Eq.~\eqref{eq:find_angles} to the block-encoded Hamiltonian $\tilde{\calW}$ (Fig.~\ref{fig:protocol}D). For an even integer $d$, the QSP unitary takes the form~\cite{Gilyen_2019,Martyn_2021_grand},
\begin{align}
\label{eq:QSP_matrix}
    &\calU_\mathrm{QSP}
    :=\prod_{k=1}^{d/2}\big[ \calS(\phi_{2k-1})\calW^\dag\calS(\phi_{2k})\calW\big]
    \nonumber\\
    &=\bigoplus_{\lambda_{\tilde{\calW}}}
    \begin{pmatrix}
        f(\lambda_{\tilde{\calW}}) & \ast \\
        \ast & \ast
    \end{pmatrix}
    \otimes \ket{\lambda_{\tilde{\calW}}}\bra{\lambda_{\tilde{\calW}}}
    =
    \begin{pmatrix}
        f(\tilde{\calW}) & \ast \\
        \ast & \ast
    \end{pmatrix},
    \\
    &\calS(\phi) := \bigoplus_{\lambda_{\tilde{\calW}}}
    \begin{pmatrix}
        \e^{\im\phi} & 0 \\
        0 & \e^{-\im\phi}
    \end{pmatrix}  
    \otimes\ket{\lambda_{\tilde{\calW}}}\bra{\lambda_{\tilde{\calW}}},
\end{align}
where the direct sum is taken over the eigenstates $\{\ket{\lambda_{\tilde{\calW}}}\}$ of $\tilde{\calW}$ and the upper-left block of the matrices represents the $\ket{0^a}\bra{0^a}$ component of the corresponding operators.
Thus, starting from the initial ancillary state $\ket{0^a}$, and post-selecting on the ancillary state $\ket{0^a}$ at the end, we obtain
\begin{align}
\label{eq:QSVT}
    (\bra{0^a}\otimes I)\calU_\mathrm{QSP}(\ket{0^a}\otimes I)
    = f(\tilde{\calW}),
\end{align}
which approximates the desired real-time evolution operator $\e^{-\im H t}$.

Let us now discuss how to post-process the measurement results and mitigate the noise effects on observables.
We let the noisy quantum state simulated on the hardware before any measurement be $\eta$, which is generally different from the state affected only by the depolarizing channel given by Eq.~\eqref{eq:noisy_state}.
For simplicity, we consider the expectation value, $\Tr[\bar{P}\eta]$, of $\bar{P}:=\ket{0^a}\bra{0^a}\otimes P$, where $P$ is a Pauli operator acting on the system register. The variance is $\mathrm{Var}_{\eta,\bar{P}}=\Tr[\bar{I}\eta] - \Tr[\bar{P}\eta]^2$.
We mitigate the noise effects by modelling it with the depolarizing channel~\cite{Arute2019,Dalzell2021,Urbanek2021}. In particular, we use the same noise model that we previously employed when estimating the optimal polynomial degree.
The expectation value of $\bar{P}$ with respect to the state in Eq.~\eqref{eq:noisy_state} is
\begin{align}
\label{eq:pauli_depol}
    \Tr[\bar{P}\calD_p[\sigma]]
    = (1-p)\bra{\psi_0}f(\tilde{\calW})^\dag P f(\tilde{\calW})\ket{\psi_0}.
\end{align}
where $p=1 - (1-p_\mathrm{TQ})^{N_\mathrm{TQ}}$.
We infer the noiseless expectation value from the noisy expectation value as
\begin{align}
\label{eq:pauli_mitig}
    \langle \bar{P}\rangle_\eta^\mathrm{mitig}
    :=\frac{\Tr[\bar{P}\eta]}{1-p}.
\end{align}
This is understood as mitigating the depolarizing noise, at the cost of a larger variance,
\begin{align}
\label{eq:var_mitigation}
    \mathrm{Var}_{\eta,\bar{P}}^{\mathrm{mitig}}
    = \frac{\mathrm{Var}_{\eta,\bar{P}}}{(1-p)^2}
    = \frac{\mathrm{Var}_{\eta,\bar{P}}}{(1-p_\mathrm{TQ})^{2N_\mathrm{TQ}}}.
\end{align}
This implies that the number of samples needed to achieve a fixed sampling error increases exponentially in $N_\mathrm{TQ}$. Therefore, reducing the depth of the circuit is extremely important even though the noise effect on the expectation value $\langle \bar{P}\rangle_\eta^\mathrm{mitig}$ is mitigated.

\subsection{Hardware experiment}
\label{sec:experiments}

In order to demonstrate the protocol, we perform the QSP-based Hamiltonian simulation experiments on the Quantinuum H1-1 trapped-ion quantum computer.
We simulate the real-time dynamics of the quantum system described by the one-dimensional Ising spin Hamiltonian
\begin{align}
\label{eq:spin_hamiltonian}
    H = -J\sum_{i=0}^{n-2}Z_{i}Z_{i+1}
    -\sum_{i=0}^{n-1} h_i X_{i}
    -m\sum_{i=0}^{n-1}Z_{i} .
\end{align}

We quantify entanglement growth by bi-partitioning the system into subsystems $A$ and $\bar{A}$ and then computing the time dependence of the von~Neumann entropy
\begin{align}
    S_\mathrm{vN} = -\Tr[\rho_A\log\rho_A],
\end{align}
and the degree-2 R\'{e}nyi entropy
\begin{align}
    S_\mathrm{R}^{(2)} = - \log\Tr[\rho_A^2] ,
\end{align}
on the $n_A$-qubit subsystem $A$, where $\rho_A=\Tr_{\bar{A}}[\rho]$.

\begin{figure}
\centering
\includegraphics[scale=0.52]{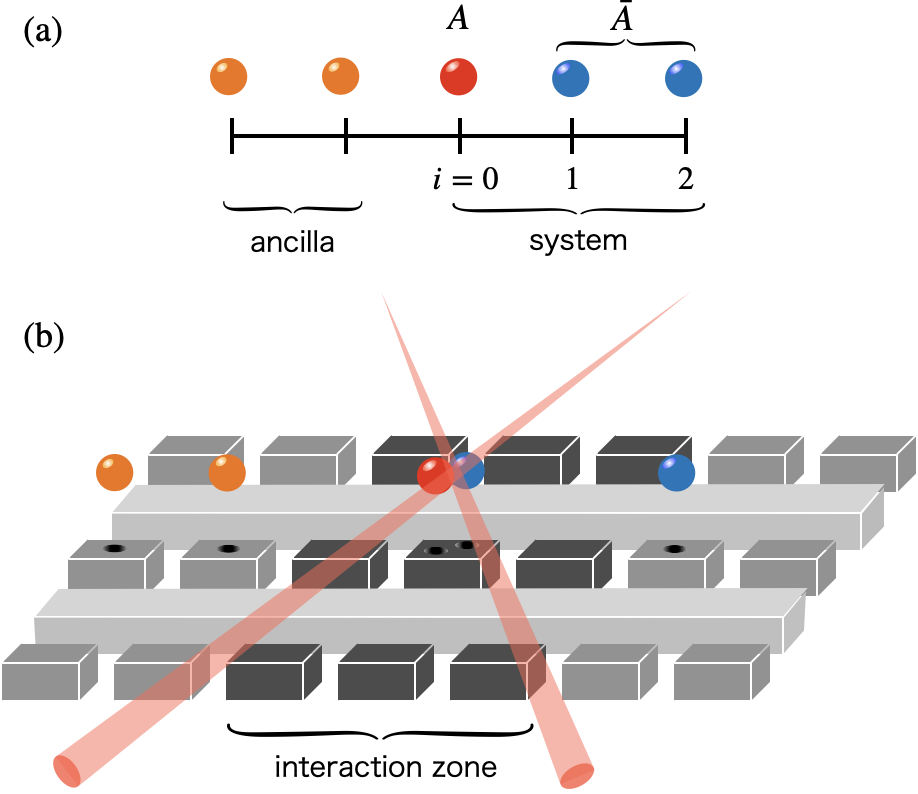}
\caption{Sketch of the setup for the five-qubit experiment. (a) The system consists of the two-qubit ancillary register (orange ions) and the three-qubit system register. The latter is further partitioned into the one-qubit subsystem $A$ (a red ion) and its complement $\bar{A}$ (blue ions).
(b) The H1-1 quantum computer operates by manipulating the ions representing the qubits. Each quantum operation (initialization, gate application, measurement) is performed using lasers after the target ions are transported to one of the isolated interaction zones. In the experiments we use five out of the 20 qubits available, and apply up to 328 two-qubit gates.}
\label{fig:H1hardware}
\end{figure}

We perform state tomography by measuring the Pauli expectation values via
\begin{align}
\label{eq:tomography_coeff}
    c_P = \frac{\langle\bar{P}\rangle_\eta^\mathrm{mitig}}{\langle\bar{I}\rangle_\eta^\mathrm{mitig}},
\end{align}
for an operator $P\in\text{Pauli}_A:=\{I,X,Y,Z\}^{\otimes n_A}\backslash\{I^{\otimes n_A}\}$ on $A$ (see Methods), which leads to an estimator of the density matrix,
\begin{align}
    \rho_A = \frac{I + \sum_{P\in\text{Pauli}_A}c_P P}{2^{n_A}}.
\end{align}

Since the denominator of Eq.~\eqref{eq:tomography_coeff} would be one in the absence of algorithmic error and noise effects, the quantity in Eq.~\eqref{eq:tomography_coeff} approximates the expectation value of the Pauli operator $P$ as is further discussed in the Methods section. 
We note that the computation of von~Neumann entropy is not scalable in general.
However, the current procedure can be straightforwardly applied to the computation of degree-2 R\'{e}nyi entropy using the swap trick~\cite{Filip2002,Ekert2002,Horodecki2002,Alves2004,Mintert2007,Johri2017} or randomized measurement protocols~\cite{Enk2012,Elben2018,Elben2019,Brydges2019,Elben2020,Elben:2022jvo}.

\begin{figure*}
\centering
\includegraphics[width=\textwidth]{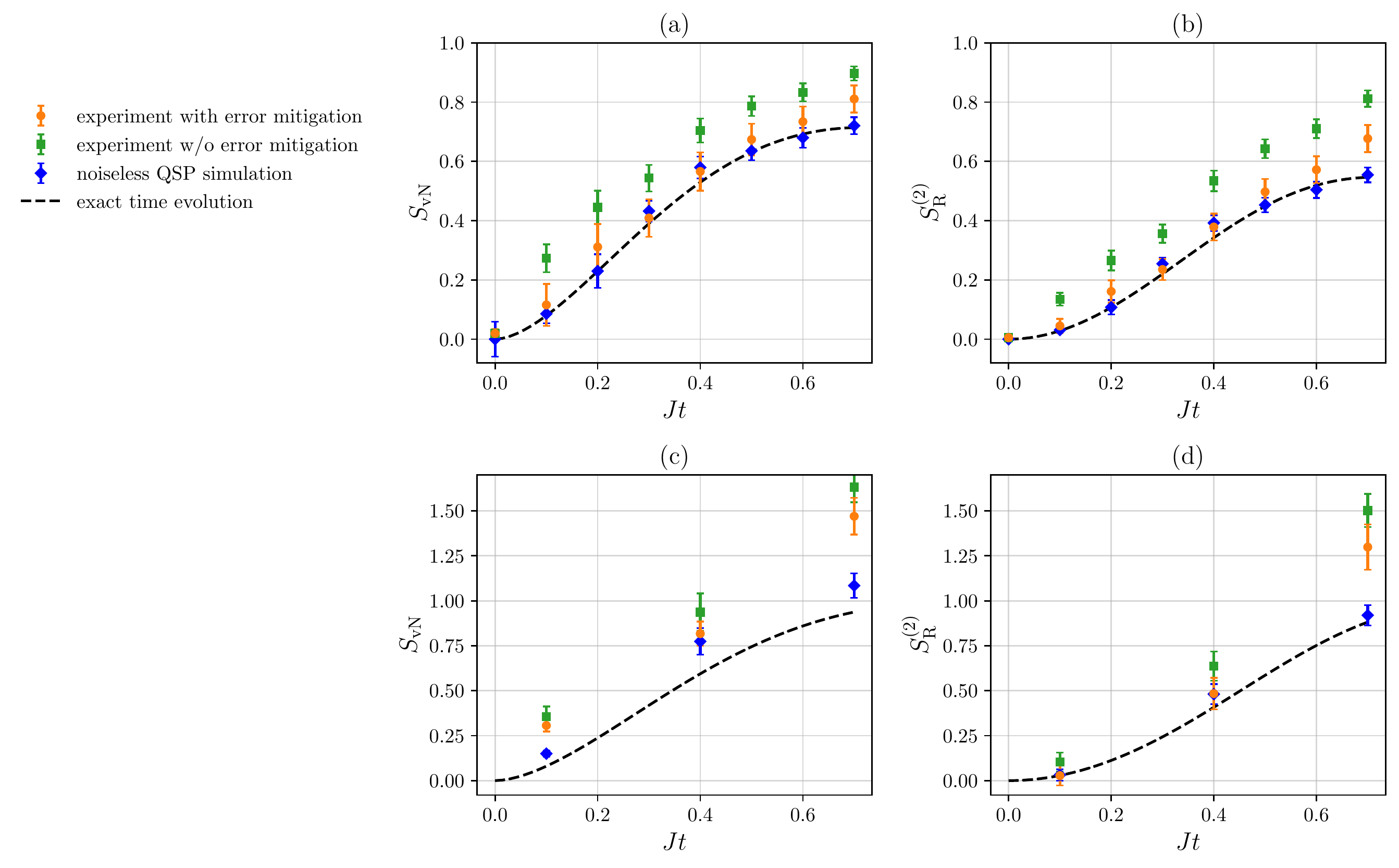}
\caption{Experimental results. (a) The von~Neumann entanglement entropy and (b) the degree-2 R\'{e}nyi entanglement entropy of the five-qubit experiment on the H1-1 quantum computer. (c) The von~Neumann entanglement entropy and (d) the degree-2 R\'{e}nyi entanglement entropy of the seven-qubit experiment. Error bars represent one standard deviation due to sampling error.}
\label{fig:entropy_H1hardware}
\end{figure*}

The H1-1 system operates by controlling the $S_{1/2}$ hyperfine clock states of trapped ${}^{171}$Yb$^+$ ions, which play the role of qubits~\cite{Kaushal2020,Pino:2020mku}; there are a total of 20 qubits in the system at the time the experiments are conducted (see \cite{H1datasheet} for details on the H1-1 system). 
In addition to single-qubit rotations, a two-qubit native gate $\exp(-\im \theta Z\otimes Z / 2)$ with $\theta\in\mathbb{R}$ can be applied to an arbitrary pair of qubits giving the system all-to-all connectivity.
This is enabled by the ability of the H1-1 system to move any pair of ions to one of five isolated interaction zones where quantum operations (initialization, gate application, measurement) are executed in a manner that suppresses the rate of crosstalk and allows for high-fidelity two-qubit gates.

In the first experiment, we consider the $n=3$ Ising spin chain with $h_i/J=-1.05$ for all $i$ and $m/J=0.5$ in Eq.~\eqref{eq:spin_hamiltonian}. The system is known to display rapid growth of entanglement~\cite{banuls_strong_2011,Shenker:2013pqa}. We preprocess the Hamiltonian $H$ given in Eq.~\eqref{eq:spin_hamiltonian} to find $\tilde{H}$ via Eq.~\eqref{eq:rescaling_H} with $a=0$, $b=1$, and $\lambda_{\pm}=\pm(2J+3h+3m)$. We obtain a compressed block-encoding circuit by variational optimization using two ancillary qubits and $L=3$ layers obtaining an error $\epsilon_\mathrm{BE}=1.8\times10^{-2}$ (see Methods for details). The subsystem $A$ is taken to be the zeroth site of the system register (see Fig.~\ref{fig:H1hardware} for a schematic of this five-qubit experiment).

\begin{table*}[ht]
\centering
\small
    \begin{tabularx}{\textwidth}{p{.14\textwidth}|p{.45\textwidth}|p{.4\textwidth}}
    Step & Description of the error & Possible improvements \\
    \hline \hline
    Preprocessing 
    &
    The spectrum of the Hamiltonian is rescaled using crude upper and lower bounds. This leads to a longer effective evolution time $\tilde{t}$ in Eq.~\eqref{eq:effective_t}. 
    & 
    Tighter bounds on the spectrum, e.g., using the methods in~\cite{Baumgratz2012lower, BaHu12, Anderson1951, Eisert2023, Kull2022}. 
    \\
    \hline
    Compressed block-encoding 
    &
    Imperfect block-encoding if variational optimization of circuit parameters is used. This leads to an error $\epsilon_\mathrm{BE}$ in Eq.~\eqref{eq:qsp_epsilon}. 
    & 
    A more expressive circuit ansatz and a higher performance classical optimizer/compiler, e.g., using the methods in~\cite{McKeever_2022}.
    \\
    \hline
    Operator-function design 
    &
    The real-time evolution operator is approximated by a polynomial of fixed degree. This leads to an error $\epsilon_\mathrm{poly}$ in Eq.~\eqref{eq:qsp_epsilon}.
    &
    A higher degree of the polynomial, e.g, using methods in~\cite{Haah2019product,Chao2020,Dong2021}. 
    \\
    \hline
    Processing 
    & 
    Each two-qubit gate fails with some probability $p_\mathrm{TQ}$. This leads to a reduced fidelity in Eq.~\eqref{eq:noisy_fidelity}. 
    &
    Quantum error detection, e.g., using the method in~\cite{Self2022protecting}.
    \end{tabularx}
    \caption{\label{tab:sources_error}
    Summary of the main sources of error in our QSP protocol and how they can be improved upon. Note, improving upon some errors affects the other errors in non-trivial ways.}
\end{table*}

We consider the real-time evolution with $Jt\in\{0,0.1,0.2,0.3,0.4,0.5,0.6,0.7\}$ and starting from the initial state on the system register 
${\ket{\psi_0}=\ket{+}^{\otimes 3}}$ where $\ket{+}=(\ket{0}+\ket{1})/\sqrt{2}$. 
For each evolution time, the degree $d$ is set to $d_\mathrm{opt}\in\{0,4,4,6,8,10,10,14\}$ following the heuristic shown in Fig.~\ref{fig:error_degree}(c). The resulting number of two-qubit gates in each circuit is $N_\mathrm{TQ} \in \{0,98,98,144,190,236,236,328\}$.
Error-mitigated Pauli expectation values in Eq.~\eqref{eq:tomography_coeff} are obtained from Eq.~\eqref{eq:pauli_mitig} with the two-qubit gate infidelity $p_\mathrm{TQ}=2.577\times10^{-3}$, the number of two-qubit gates $N_\mathrm{TQ}$, and 1000 measurements.

Figures~\ref{fig:entropy_H1hardware}(a) and (b) show the growth of entanglement entropies with time for our system.
The exact time evolution data (dashed line) is obtained from the exact application of the operator $\e^{-\im H t}$ to the initial state $\ket{\psi_0}$. 
The experimental data obtained from H1-1 is reported with error mitigation (orange circles) as well as without error mitigation (green squares). 
The noiseless QSP simulation data (blue diamonds) is obtained by classically simulating the algorithm without the noise effects. 
Error bars represent one standard deviation due to sampling error.

The error-mitigated experimental data agree well with the exact values and with the noiseless data up to $Jt=0.6$, while there is a discrepancy between the unmitigated data and the rest from as early as $Jt=0.1$. We also observe that the error-mitigated data show larger sampling errors (error bars) than the unmitigated data as expected from Eq.~\eqref{eq:var_mitigation}.
The experimentally obtained entanglement entropies generally yield larger values than the exact ones due to algorithmic error and noise effects, which induce the interaction among the system register, ancillary register, and environment surrounding the device. Thus, the von~Neumann and R\'{e}nyi entropies computed on the subsystem $A$ measure the entanglement not only with the system $\bar{A}$ but also with the ancillary register and environment.
Nevertheless, our protocol mitigates these erroneous impacts well. In particular, the agreement between the mitigated experimental data and exact values indicates that our protocol brings both QSP algorithmic error and noise effects under good control for the range of parameters that we assessed.

In the second experiment, we simulate the real-time evolution of the $n=4$ Ising spin chain with $h_1/J=1$ and $h_i/J=m/J=0$ for $i\neq1$ in Eq.~\eqref{eq:spin_hamiltonian}.
We begin by constructing the exact LCU block-encoding circuit ($\epsilon_\mathrm{BE}=0$) which uses $a=3$ ancillary qubits and 125 two-qubit gates. We compress this circuit using multiplexor compilation and obtain an equivalent circuit with only 44 two-qubit gates. This is a reduction of $64.8\%$ of the original LCU circuit size (see Methods for details).
We evolve the initial state ${\ket{\psi_0}=\ket{+}^{\otimes 4}}$ on the system register and make 1000 measurements to compute each Pauli expectation value~[Eq.~\eqref{eq:tomography_coeff}] at each time $Jt\in\{0.1,0.4,0.7\}$. 
We again follow the heuristic in Fig.~\ref{fig:protocol}C to find $d_\mathrm{opt}\in\{2,4,8\}$ for each evolution time $Jt$. However, we use a different two-qubit gate infidelity, $p_\mathrm{TQ}=2.185\times10^{-3}$, following an update to the H1-1 device after our first experiment. The resulting number of two-qubit gates in each circuit is $N_\mathrm{TQ} \in \{102, 204, 408\}$.

We choose the zeroth and first sites of the system register to represent subsystem $A$. The calculated entanglement entropies are shown in Figs.~\ref{fig:entropy_H1hardware}(c) and (d). The discrepancy between the noiseless data (blue diamonds) and exact data (dashed line) is due to the degrees $d_\text{opt}$ being smaller than those found in the first experiment. Indeed, the heuristic has taken into account the increased number of qubits and two-qubit gates for this second experiment. The degrees found by our heuristic lead to a good agreement between the noiseless data and error-mitigated experimental data (orange circles), except for $Jt=0.7$. Note that this parameter setting ($Jt=0.7$) yields our largest quantum circuit with as many as 408 two-qubit gates. This experiment exemplifies the importance of finding the optimal working point to balance the algorithmic error, hardware noise, and parameter setting.

\section{Discussion}
\label{sec:conclusion}

We propose a detailed protocol to perform QSP-based Hamiltonian simulation tailored to noisy quantum hardware.
Each process is carefully studied to clarify the sources of error in the estimate of target observables, as summarized in Tab.~\ref{tab:sources_error}.
In particular, the polynomial approximation is designed such that the combined error caused by the QSP protocol and noise effect is minimized.
The block-encoding circuit is compressed to further reduce the circuit depth for experimental purposes. 
An error mitigation scheme is used to increase accuracy in the estimate of target expectation values.

We execute the protocol on the Quantinuum H1-1 quantum computer.
As an illustration, the time evolution of von~Neumann and degree-2 R\'{e}nyi entanglement entropies are computed. The results from the hardware experiments agree not only with those from noiseless simulations but with exactly obtained values, which implies the algorithmic error and noise effects are well controlled in the range of parameters that we chose.

An important question is whether the approach can scale to larger demonstrations. Both our heuristic and error mitigation schemes are derived under a simple noise model for the hardware at hand. A sophisticated error model may be required to obtain more accurate outputs for larger instances. Beyond that, one can use quantum error detection codes (see, e.g., \cite{Self2022protecting} for the code tailored for the Quantinuum H1 system) to generate more reliable results at the cost of discarding a portion of the circuit runs, or apply algorithm-level error correction~\cite{Tan:2023nwy} for noisy QSP. 
Finally, it is noted that there exist block-encoding schemes with asymptotically efficient scaling~\cite{Gilyen_2019,Low2019hamiltonian,Chakraborty2019power,Camps:2022jnx,Camps:2022uyk}. Their required quantum resources are, however, still beyond the capability of currently available quantum devices. The techniques employed in this article to compress block-encoding circuits are potentially useful to perform larger-scale QSP realizations.

While further theoretical improvements are still required to scale up the protocol, the present study has taken the first step in the experimental realization of QSP-based algorithms and applications. 

\section{Methods}

\subsection{Compressed block-encoding by variational optimization}
\label{app:compressBE}

Here we elaborate on the block-encoding techniques used in this work. 
The goal is to optimize a parameterized quantum circuit, $\calW(\bm{\theta})$,
to minimize the block-encoding error,
\begin{align}
    &\epsilon_\mathrm{BE} 
    = \lVert\tilde{\calW}(\bm{\theta})-\tilde{H}\rVert_\mathrm{F},
    \\
    &\tilde{\calW}(\bm{\theta}) = (\bra{0^a}\otimes I^{\otimes n})\calW(\bm{\theta})(\ket{0^a}\otimes I^{\otimes n}),
\end{align}
with $\bm{\theta}$ referring to the collection of all the parameters in the circuit.
This is equivalent to minimizing the cost function,
\begin{align}
\label{eq:cost_fn}
	F(\bm{\theta}) = \Tr(\tilde{\calW}^\dag\tilde{\calW}) - 2\mathrm{Re}\Tr(\tilde{H}\tilde{\calW}),
\end{align}
where we used that $\tilde{H}$ is a Hermitian operator.
Provided that the Hamiltonian is expanded as $\tilde{H}=\sum_\ell c_\ell P_\ell$ with $n$-qubit Pauli operators $\{P_\ell\}$, the error $\epsilon_\mathrm{BE}$ is obtained from $F(\bm{\theta})$ by
\begin{align}
\label{eq:frob_dist}
    (\epsilon_\mathrm{BE})^2 = F(\bm{\theta})-\Tr(\tilde{H}^2)
    = F(\bm{\theta})-2^n \sum_\ell c_\ell^2.
\end{align}

\begin{figure*}
    \centering
    \includegraphics[width=\textwidth]{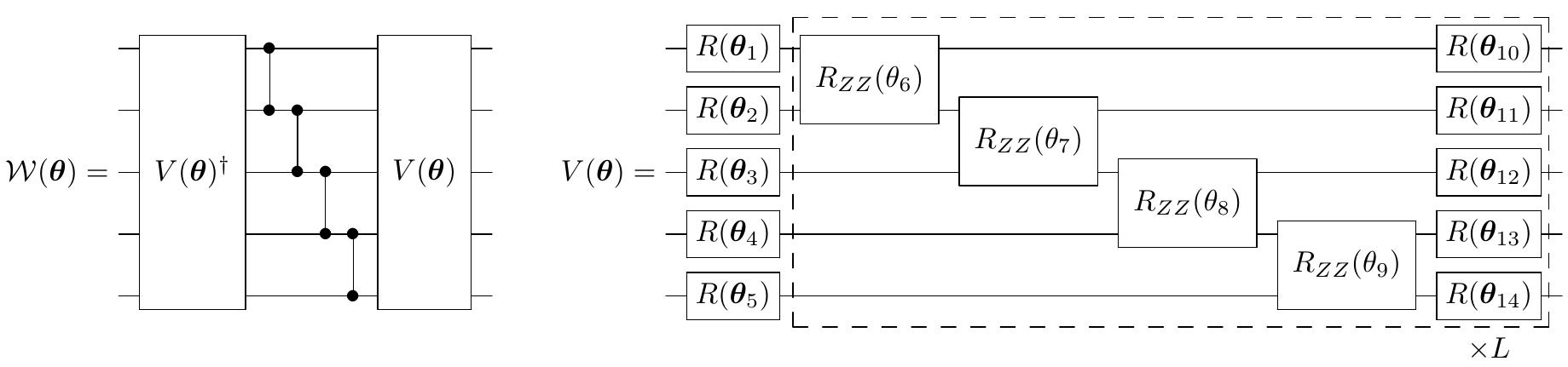}
    \caption{Quantum circuit diagrams for compressed block-encoding by variational optimization. (Left) An example of $(a+n)$-qubit parameterized quantum circuit $\calW(\bm{\theta})$ satisfying the qubitization condition. (Right) An example of sub-circuit $V(\bm{\theta})$. The circuit inside the dashed box is repeated $L$ times with new variational parameters added for each layer. 
    The single- and two-qubit gates used in the circuit are $R(\bm{\theta})=\exp(-\im\theta^{(3)}X/2)\exp(-\im\theta^{(2)}Z/2)\exp(-\im\theta^{(1)}X/2)$ and $R_{ZZ}(\theta)=\exp(-\im\theta Z\otimes Z/2)$. In our five-qubit experiment, we use the bottom $n(=3)$ qubits as the system register and the top $a(=2)$ qubits as the ancillary register.}
    \label{fig:ansatz}
\end{figure*}

We consider a particular structure for the parameterized quantum circuit which satisfies the reflection condition $\calW(\bm{\theta})^2=I^{\otimes n}$. This condition is not crucial to the construction of QSP. However, we empirically found that the constraint makes optimization of block encoding easier. 
One ansatz satisfying the reflection condition is shown in Fig.~\ref{fig:ansatz} and given by
\begin{align}
\label{eq:ansatz}
	\calW(\bm{\theta}) = V(\bm{\theta})\,\overline{CZ}\,V(\bm{\theta})^\dag,
\end{align}
where $V(\bm{\theta})$ is a unitary operator specified by the right circuit of Fig.~\ref{fig:ansatz}, and $\overline{CZ}$ stands for the sequential application of controlled-$Z$ gates that is shown in the middle of the upper circuit.

The parameterized quantum circuit $\calW(\bm{\theta})$ shown in Fig.~\ref{fig:ansatz} is composed of the following gates:
\begin{align}
\label{eq:NativeGates}
    &R_{X}(\theta) = \exp(-\im \theta X / 2), \\
    &R_{Z}(\theta) = \exp(-\im \theta Z / 2), \\
    &R_{ZZ}(\theta) = \exp(-\im \theta Z \otimes Z / 2),
\end{align}
where each gate has an independent variational parameter $\theta$.
Importantly, these gates are part of the native gate set of the Quantinuum H1-1 quantum computer.

\begin{figure*}
\centering
\includegraphics[width=.85\textwidth]{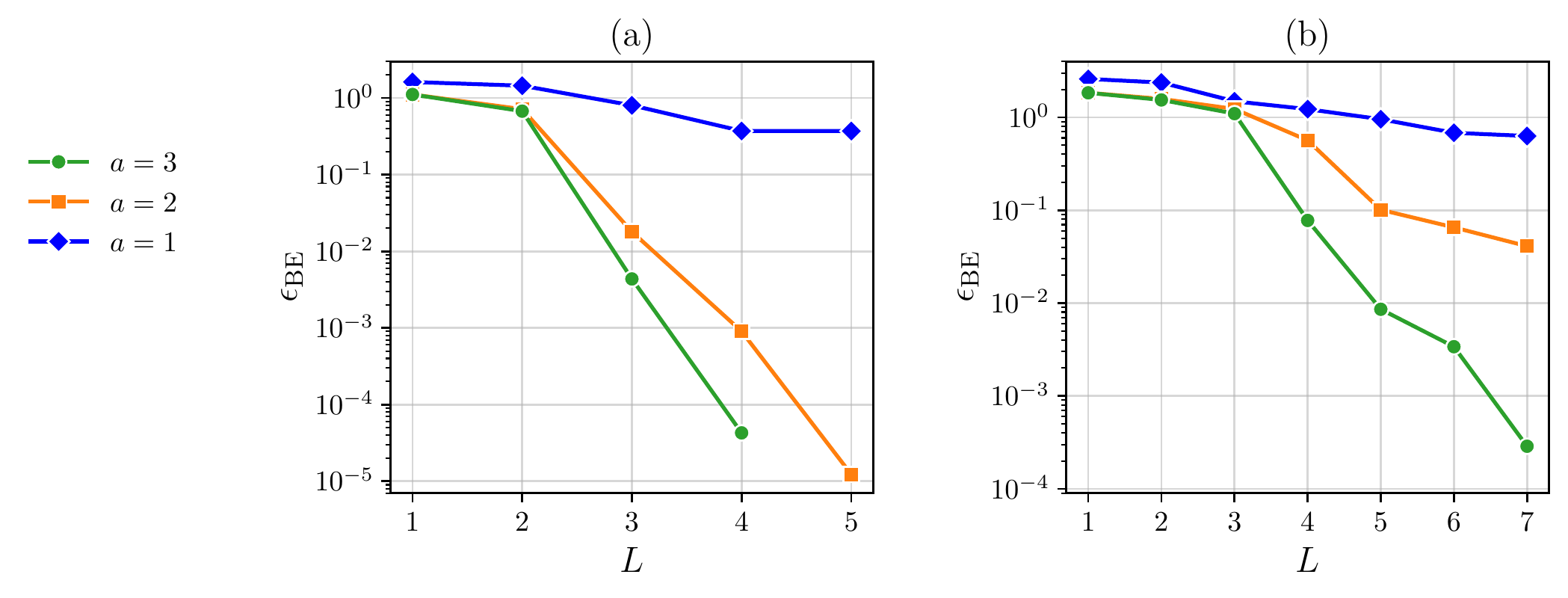}
\caption{Error $\epsilon_\mathrm{BE}$ of the block encoding circuit as a function of the number of layers $L$ and for each number of ancillary qubits~$a$. We use the Ising spin Hamiltonian with $h_i/J=-1.05$ for all $i$ and $m/J=0.5$. The system size $n$ is three in (a) and four in (b).}
\label{fig:varBE}
\end{figure*}

In the present work, the optimization of the block-encoding circuit is performed by minimizing the cost function given in Eq.~\eqref{eq:cost_fn} using a classical state-vector simulation and the quasi-Newton BFGS method~\cite{Nocedal2009}.
The optimization is stopped when the gradient norm of the cost function falls below the threshold value $1\times10^{-5}$.
The accuracies of the optimized block encoding circuits for the 3-site and 4-site Ising spin Hamiltonian are shown in Fig.~\ref{fig:varBE}.
In the experiment of the 3-site Ising spin chain, we use the circuit with $a=2$ and $L=3$, which requires $(a+n-1)(2L+1)=28$ $R_{ZZ}$ gates.
The optimized circuit has block-encoding error $\epsilon_\mathrm{BE}=1.8\times10^{-2}$.

\begin{figure*}
\centering
\includegraphics[width=123.85mm]{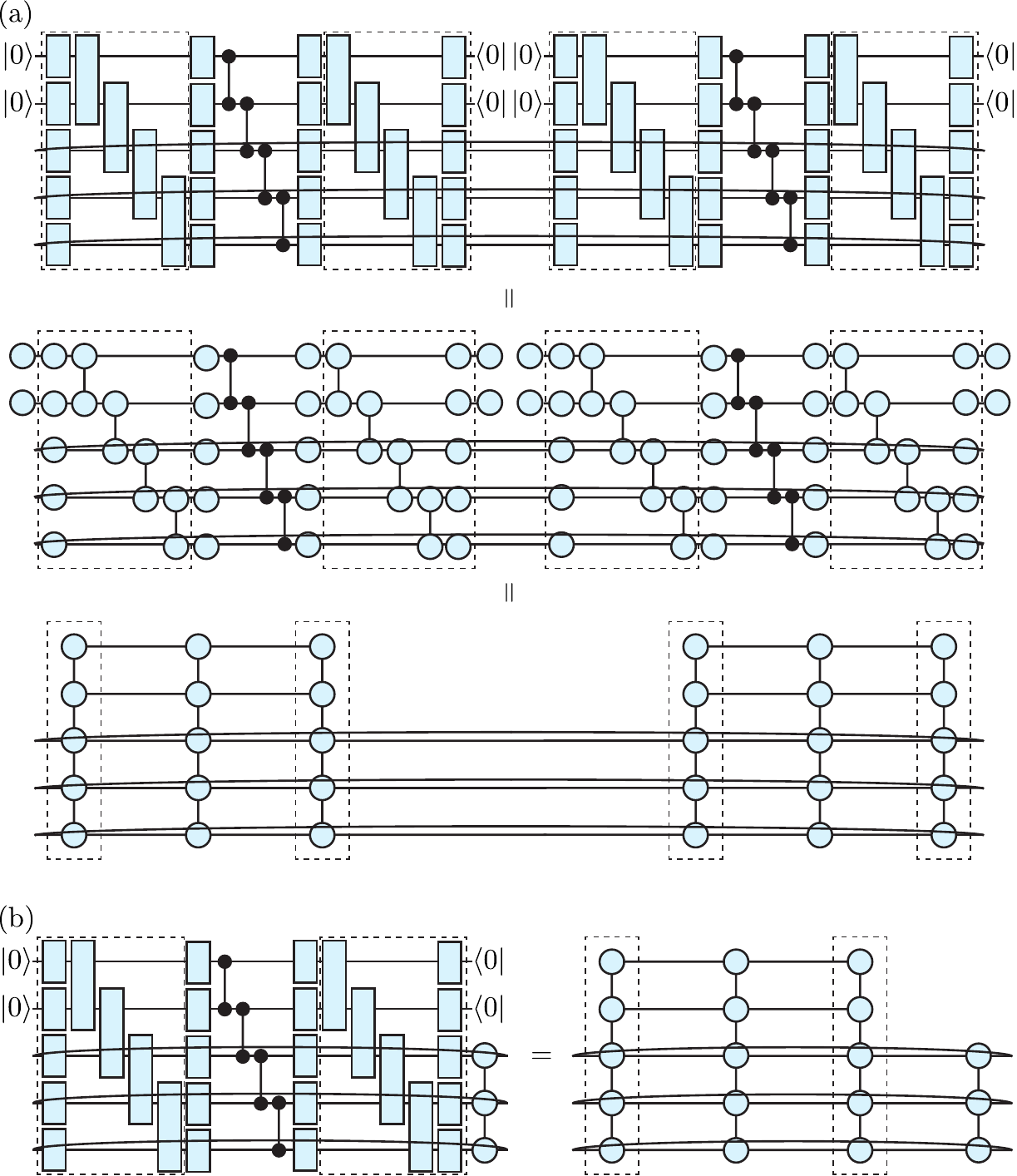}
\caption{Tensor network contractions for the evaluation of the cost function. (a) Contraction of $\Tr(\tilde{\calW}^{\dag} \tilde{\calW})$ for $\tilde{\calW}$ of Fig.~\ref{fig:ansatz}(a).
(b) Contraction of $\Tr(\tilde{\calW}^{\dag}\tilde{H})$ for $\tilde{\calW}$ of Fig.~\ref{fig:ansatz}(a) and $\tilde{H}$ represented by a matrix product operator.
Note that the terms in the gradient~\eqref{eq:gradientIter} and Hessian~\eqref{eq:hessian} can be evaluated using similar tensor network contractions.}
\label{fig:Contraction}
\end{figure*}

We briefly discuss a classical method based on tensor network techniques.
By expressing the cost function [Eq.~\eqref{eq:cost_fn}] as a tensor network contraction and using a classical optimizer to find the parameters $\boldsymbol{\theta}$, a block-encoding circuit $\mathcal{W}(\boldsymbol{\theta})$ which minimizes $\epsilon_\mathrm{BE}$ can be found.
The terms in the cost function Eq.~\eqref{eq:cost_fn}, $\Tr(\tilde{\calW}^\dag\tilde{\calW})$ and $\Tr(\tilde{H}\tilde{\calW})$, can be evaluated using tensor network contractions as illustrated in Fig.~\ref{fig:Contraction}.

The cost function in Eq.~\eqref{eq:cost_fn} can be variationally optimized using a classical optimizer, for instance, we can employ a gradient-based method as follows. At each iteration $i$, we require the gradient vector $\bm{\mathcal{G}}^{(i)}$ of the objective function $F(\bm{\theta})$ at $\bm{\theta}=\bm{\theta}^{(i)}$:
\begin{align}
    \mathcal{G}_{k}^{(i)}
    & = \frac{\partial F}{\partial \theta_{k}}
    = 2\mathrm{Re}\left[\Tr\left(\tilde{\calW}^{\dag} \frac{\partial\tilde{\calW}}{\partial \theta_k}\right)\right] 
    - 2\mathrm{Re}\left[\Tr\left(\tilde{H}\frac{\partial\tilde{\calW}}{\partial \theta_k}\right)\right].
\end{align}

The partial derivatives in each gradient are straightforward to compute via the first of the variational gates given in Eq.~\eqref{eq:NativeGates}.
We then iterate
\begin{equation}
\label{eq:gradientIter}
    \boldsymbol{\theta}^{(i+1)} = \boldsymbol{\theta}^{(i)} - \gamma\,\boldsymbol{\mathcal{G}}^{(i)},
\end{equation}
with some learning parameter $\gamma>0$ to update the parameters. The iteration is repeated until the norm of the vector of gradients falls below a predefined convergence threshold.

One could improve the convergence rate by additionally computing the Hessian matrix $\mathcal{H}^{(i)}$ at the cost of more evaluations of operator expectation values:
\begin{align}
\label{eq:hessian}
    &\mathcal{H}_{j, k}^{(i)}
    = \frac{\partial^{2} F}{\partial \theta_{j} \partial \theta_{k}}
    = 2\mathrm{Re}\left[\Tr\left(\tilde{\calW}^{\dag} \frac{\partial^2\tilde{\calW}}{\partial \theta_j\partial \theta_k}\right)\right] \\
    &\qquad +2\Tr\left(\frac{\partial\tilde{\calW}^{\dag}}{\partial\theta_j} \frac{\partial\tilde{\calW}}{\partial \theta_k}\right) -2\mathrm{Re}\left[\Tr\left(\tilde{H}\frac{\partial^2\tilde{\calW}}{\partial \theta_j\partial \theta_k}\right)\right].
\end{align}
Then, the parameter update in Eq.~\eqref{eq:gradientIter} is replaced with,
\begin{equation}
\label{eq:NewtonIter}
    \boldsymbol{\theta}^{(i+1)} = \boldsymbol{\theta}^{(i)} - \left( \mathcal{H}^{(i)} \right)^{-1} \boldsymbol{\mathcal{G}}^{(i)}.
\end{equation}
For the computation of the inverse of the Hessian matrix, we use the fact that this matrix is Hermitian and since our goal is to minimize the objective function in Eq.~\eqref{eq:cost_fn}, we are only interested in its positive eigenvalues.

Therefore we compute the pseudo-inverse via the eigendecomposition of the Hessian matrix and set all eigenvalues $\mu_{k}$ smaller than some small cutoff $\epsilon$ to zero, e.g., \ $\epsilon = 1\times 10^{-5}$.
More specifically, the pseudo-inverse is computed by replacing $\mu_{k}$ by $1/\mu_{k}$ in the diagonal matrix of the eigendecomposition using only the positive eigenvalues $\mu_{k} \geq \epsilon$ (all other eigenvalues are set to zero).

\subsection{Compressed block-encoding by multiplexor compilation}

As an alternative approach to compressing a block-encoding circuit, 
we employ the linear-combination-of-unitaries (LCU) method~\cite{Childs2012} with the help of an efficient compilation of multi-controlled unitary gates (multiplexors). 
LCU provides a way to block encode $\tilde{H}$ when it is expressed as a weighted sum of unitary operators, $\{P_\ell\}_{\ell=1}^{K}$, $\tilde{H}=\sum_{\ell=1}^K c_\ell P_\ell$.
The LCU consists of two unitary operators: 
\begin{enumerate}
   \item[i)] an operator $A$ acting on the ancillary register with $a=\lceil \log_2 K \rceil$ such that $A\ket{0^a} = \frac{1}{\sqrt{c}}\sum_{\ell=1}^K \sqrt{c_\ell}\ket{\ell}$ with $c=\sum_{\ell=1}^K c_\ell$; and
   \item[ii)] a controlled operator $B = \sum_{\ell=1}^{K}\mathrm{sign}(c_\ell)\ket{\ell}\bra{\ell}\otimes P_\ell$ with the sign function, $\mathrm{sign}(c)=+1(-1)$ for $c\ge0(c<0)$.
\end{enumerate}
With these, 
\begin{align}
\label{eq:LCU_block_enc}
    \mathcal{W}=A^\dag BA
\end{align}
gives an exact block encoding of $\tilde{H}$, i.e., $\epsilon_\mathrm{BE}=0$.

\begin{figure*}
    \centering
    \includegraphics[width=.68\textwidth]{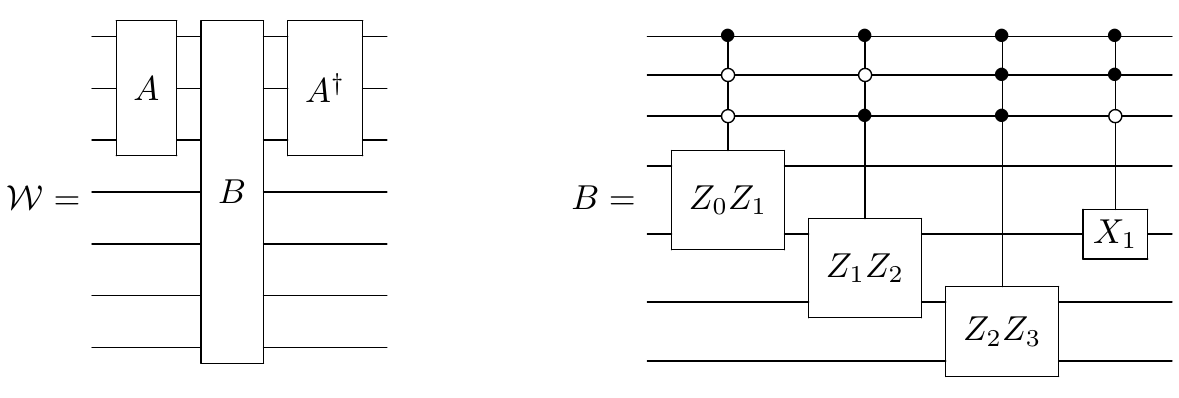}
    \caption{Quantum circuit diagrams for compressed block-encoding by multiplexor compilation. (Left) Structure of the LCU-based block encoding $\calW$ given by Eq.~\eqref{eq:LCU_block_enc}. The top three and bottom four qubits represent the ancillary and system registers, respectively. (Right) The sub-circuit $B$ used for block-encoding the $n=4$ Ising spin Hamiltonian with $h_1/J=1$ and $h_i/J=m/J=0$ for $i\neq1$, before the multiplexor compilation is applied.}
    \label{fig:LCU}
\end{figure*}

The bottleneck of this construction is the implementation of $B$, which contains a sequential application of multi-controlled-$P_\ell$ gates.
We make use of the compilation technique of multiplexor, which is developed in~\cite{Yao_to_appear} based on \cite{Shende2006,Bergholm2005}, to reduce the gate complexity without introducing extra ancillary qubits.
In the block-encoding of $\tilde{H}$, we use $A=\mathrm{Had}^{\otimes 3}$ with the Hadamard gate, $\mathrm{Had}$, and apply the multiplexor compilation to $B$ shown in the right panel of Fig.~\ref{fig:LCU}.
This results in 44 $R_{ZZ}$ gates for the block-encoding circuit $\calW$. Indeed, the number of $R_{ZZ}$ gates is significantly reduced relative to the circuit obtained without the compilation, which uses 125 $R_{ZZ}$ gates.

\subsection{Heuristic estimation of the optimal degree}
\label{app:opt_degree}

One key aspect of this work is the estimation of the optimal degree for the QSP polynomial given a certain noise rate. Our heuristic uses the upper bound $\epsilon_\mathrm{total}$ on the infidelity between the noisy and target states under a simplified noise model. Here we discuss the noise model and provide further numerical results.

For our numerical study, we replace all the two-qubit gates, $R_{ZZ}(\theta)=\exp(-\im\theta Z\otimes Z/2)$ for $\theta\in\mathbb{R}$, by two-qubit depolarizing channels:
\begin{align}
\label{eq:TQ_depol}
    R_{ZZ}(\theta)\sigma R_{ZZ}(\theta)^\dag
    &\mapsto
    (1-p_2)R_{ZZ}(\theta)\sigma R_{ZZ}(\theta)^\dag \\
    &+ \frac{p_2}{15}\sum_{P\in\{I,X,Y,Z\}^{\otimes2}\backslash \{I^{\otimes2}\}}P\sigma P,
\end{align}
where $\sigma$ is some quantum state and we use the error parameter $p_2=2.416\times10^{-3}$. This value is the two-qubit fault probability reported in the System Model H1 Emulator Product Data Sheet~\cite{H1datasheet}. 
In particular, in the System Model H1-1 Emulator, the probability $p_2$ is chosen such that the faulty $R_{ZZ}(\pi/2)$ modeled by the following two-qubit depolarizing channel $D^{(2)}$ combined with the other noise channels emulates the noise of Quantinuum H1-1 quantum computer:
\begin{align}
\begin{split}
    D^{(2)}[\sigma]
    &=(1-p_2)R_{ZZ}(\pi/2)\sigma R_{ZZ}(\pi/2)^\dag \\
    &\qquad + \frac{p_2}{15}\sum_{P\in\{I,X,Y,Z\}^{\otimes2}\backslash\{I^{\otimes2}\}}P\sigma P
    \\
    &=\left(1-\frac{16p_2}{15}\right)\sigma
    + \frac{16p_2}{15}
    \Tr^{(2)}[\sigma]\otimes\frac{I^{\otimes 2}}{4},
\end{split}
\end{align}
where $\Tr^{(2)}$ indicates the trace over the two-dimensional subspace which the channel $D^{(2)}$ acts on.
We remark that, in the H1-1 Emulator, the faulty $R_{ZZ}(\theta)$ is modeled by the channel $D^{(2)}$ with $\theta$-dependent fault probability $p_2(\theta)$ (see \cite{H1datasheet} for more details).
In the present work, we simplify the noise model by using $p_2=2.416\times10^{-3}$ for all the two-qubit gates, $R_{ZZ}(\theta)$, independent of the angle $\theta$ as given by Eq.~\eqref{eq:TQ_depol}.
To clarify the relation between this parameter and the error parameter $p_\mathrm{TQ}$ used throughout our protocol (see Figure~\ref{fig:protocol}), we note that the same channel $D^{(2)}$ is expressed as
\begin{align}
    D^{(2)}[\sigma]=\left(1-p_\mathrm{TQ}\right)\sigma
    + p_\mathrm{TQ}\Tr^{(2)}[\sigma]\otimes\frac{I^{\otimes 2}}{4}.
\end{align}
Therefore, the new error parameter is identified with $p_\mathrm{TQ} = (16/15)p_2=(16/15)2.416\times10^{-3}=2.577\times10^{-3}$. This is the error parameter used in our infidelity bound.

\begin{figure*}
\centering
\includegraphics[width=.85\textwidth]{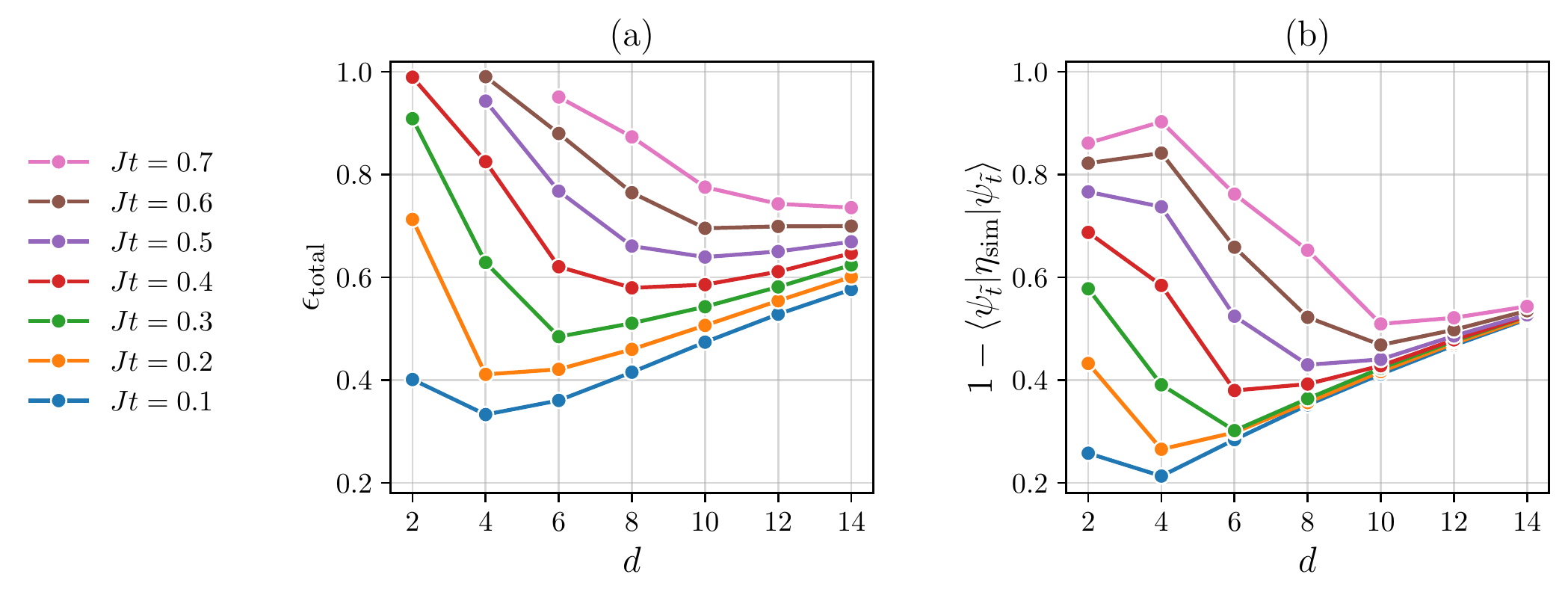}
\caption{Numerical verification of the inﬁdelity bound used in this work. (a) The upper bound of the infidelity between the target and simulated states. (b) The infidelity between the target state and simulated state with the noise model in Eq.~\eqref{eq:TQ_depol}. The locations of minima in (a) and (b) are close to each other for each time $Jt$.}
\label{fig:compare_error}
\end{figure*}

To strengthen our argument, we verify the infidelity bound using exact density matrix emulations of noisy quantum circuits. 
We let the density matrix numerically obtained by the QSP protocol with the noise channel~\eqref{eq:TQ_depol} be $\eta_\mathrm{sim}$. 
Figure~\ref{fig:compare_error}(a) shows the infidelity bound, while (b) shows the exact infidelity.
It is seen that the locations of minima in Figs.~\ref{fig:compare_error}(a) and (b) are close to each other for each evolution time $Jt$.
This observation supports that the degree $d$ minimizing $\epsilon_\mathrm{total}$ is likely to lead to the smallest possible error on noisy hardware. We emphasize that our heuristic does not require the use of a quantum computer beforehand. The optimal degree is found numerically using classical computation.

\subsection{Processing with depolarizing error mitigation}
\label{app:normalization}

In our hardware experiment we employed state tomography to compute the entanglement entropies. To this end, we estimated the expectation value of a Pauli operator $P$ on the system register by
\begin{align}
\label{eq:mitig_exp_app}
    \frac{\langle\bar{P}\rangle_\eta^\mathrm{mitig}}{\langle\bar{I}\rangle_\eta^\mathrm{mitig}}.
\end{align}
This is understood as taking the expectation of $P$ with the normalized post-selected state.
Given an initial quantum state $\ket{\psi_0}$ on the system register, we wish to approximate the time-evolved state $\e^{-\im Ht}\ket{\psi_0}\bra{\psi_0}\e^{\im Ht}$ by applying the QSP unitary
\begin{align}
    \sigma=\calU_\mathrm{QSP}(\ket{0^a}\bra{0^a}\otimes\ket{\psi_0}\bra{\psi_0})\calU_\mathrm{QSP}^\dag,
\end{align}
followed by the post-selection.
We simulate the protocol on the quantum hardware. Let $\eta$ be the experimentally obtained state on the system and ancillary registers before any measurements, and let $\tilde{\eta}$ be the state that is post-selected on the ancillary state $\ket{0^a}$ and normalized,
\begin{align}
    \tilde{\eta} 
    = \frac{(\bra{0^a}\otimes I^{\otimes n})\eta(\ket{0^a}\otimes I^{\otimes n})}{\Tr[(\ket{0^a}\bra{0^a}\otimes I^{\otimes n})\eta]}.
\end{align}
Then, the expectation value of a Pauli operator $P$ with respect to $\tilde{\eta}$ is
\begin{align}
\label{eq:pauli_mean}
    \langle P\rangle_{\tilde{\eta}}
    = \frac{\Tr[(\ket{0^a}\bra{0^a}\otimes P)\eta]}{\Tr[(\ket{0^a}\bra{0^a}\otimes I)\eta]}
    = \frac{\langle\bar{P}\rangle_\eta}{\langle\bar{I}\rangle_\eta},
\end{align}
This can be estimated with $n_\mathrm{shots}$ circuit executions with the variance
\begin{align}
    \mathrm{Var}_{\tilde{\eta},P} 
    = \langle P\rangle_{\tilde{\eta}}^2\left(\frac{\mathrm{Var}_{\eta,\bar{P}}}{\langle\bar{P}\rangle_\eta^2}
    + \frac{\mathrm{Var}_{\eta,\bar{I}}}{\langle\bar{I}\rangle_\eta^2}\right),
\end{align}
where the variances inside the parenthesis are given by $\mathrm{Var}_{\eta,\bar{P}}=(\langle\bar{I}\rangle_\eta - \langle\bar{P}\rangle_\eta^2)/(n_\mathrm{shots}-1)$ and $\mathrm{Var}_{\eta,\bar{I}}=(\langle\bar{I}\rangle_\eta - \langle\bar{I}\rangle_\eta^2)/(n_\mathrm{shots}-1)$.

To mitigate noise effects, we model them by a depolarizing channel $D_p$~\cite{Urbanek2021} applied to the entire system. Upon application of $D_p$, the state $\sigma$ becomes
\begin{align}
    D_p[\sigma] = (1-p)\sigma + p\frac{I^{\otimes n+a}}{2^{n+a}},
\end{align}
where $p = 1-(1-p_{\mathrm{TQ}})^{N_\mathrm{TQ}}$ with $N_\mathrm{TQ}$ two-qubit gates of gate infidelity $p_\mathrm{TQ}$.
With the state $D_p[\sigma]$, the expectation values of $\bar{P}$ and $\bar{I}$ take forms,
\begin{align}
    &\langle\bar{P}\rangle_{D[\sigma]}
    = (1-p)\langle\bar{P}\rangle_{\sigma},
    \\
    &\langle\bar{I}\rangle_{D[\sigma]}
    = (1-p)\langle\bar{I}\rangle_{\sigma} + \frac{p}{2^a}.
\end{align}
Thus, inverting these equations leads to the expectation values without the depolarizing noise, $\langle\bar{P}\rangle_{\sigma}=\langle\bar{P}\rangle_{D[\sigma]}/(1-p)$ and $\langle\bar{I}\rangle_{\sigma}=(\langle\bar{I}\rangle_{D[\sigma]}-p/2^a)/(1-p)$.
Assuming that the dominant source of error in the experimentally obtained state $\eta$ is depolarizing noise, we infer the noiseless expectation value as,
\begin{align}
    \langle P\rangle_{\tilde{\eta}}^{\mathrm{mitig}}
    = \frac{\langle\bar{P}\rangle_\eta^{\mathrm{mitig}}}{\langle\bar{I}\rangle_\eta^{\mathrm{mitig}}}
    = \frac{\langle\bar{P}\rangle_\eta}{\langle\bar{I}\rangle_\eta-p/2^a}.
\end{align}
This is Eq.~\eqref{eq:mitig_exp_app} and is understood as mitigating the depolarizing noise, at the cost of a larger variance,
\begin{align}
    \mathrm{Var}_{\tilde{\eta},P}^{\mathrm{mitig}}
    = \langle P\rangle_{\tilde{\eta}}^2\left(\frac{\mathrm{Var}_{\eta,\bar{P}}}{\langle\bar{P}\rangle_{\eta}^2}
    + \frac{\mathrm{Var}_{\eta,\bar{I}}}{(\langle\bar{I}\rangle_{\eta}-p/2^a)^2}\right).
\end{align}

Note that the quantity in the denominator of the second term evaluates to
\begin{align}
    \langle\bar{I}\rangle_{\eta}-\frac{p}{2^a}
    \approx
    (1-p)+\frac{p}{2^a} - \frac{p}{2^a}
    =(1-p_\mathrm{TQ})^{N_\mathrm{TQ}},
\end{align}
where the approximate equality is due to the QSP algorithmic error and other types of noise effects.
This implies that the variance, and hence the required number of samples, increases exponentially in $N_\mathrm{TQ}$ to achieve some fixed sampling error.

\section{Data availability}
The data that support the findings of this study are available at Zenodo~\cite{kikuchi_yuta_2023_8313653}.

\section{Code availability}
The code used to create the figures in this paper is available from the authors upon reasonable request.

\section{Acknowledgements} 
We thank Silas Dilkes, Samuel Duffield, Megan Kohagen, Kirill Plekhanov, Ciar\'{a}n Ryan-Anderson, Yao Tang, Oscar Watts, and Kentaro Yamamoto for helpful discussions. We thank Nathan Fitzpatrick and Matthias Rosenkranz for providing feedback on an earlier version of this paper.

\section{Author contributions}
Y.K., M.L. and M.B. conceived and designed the study. Y.K. performed analytic calculations, and Y.K. and M.B. carried out numerical studies. All authors analysed the data, created the figures, interpreted the results and wrote the manuscript.

\section{Competing interests}
The authors declare no competing financial or non-financial interests.

\end{document}